\documentstyle[12pt]{article}

\begin{document}
\title{Update of the electromagnetic effective coupling constant 
$\alpha(M^2_Z)$}
\author{N.V.Krasnikov\\ INR RAS, Moscow 117312 \\ and \\
R.Rodenberg\\
RWTH Aachen, III. Physikalisches Institut \\ Abteilung f\"ur 
Theoretische Elementarteilchenphysik\\ Physikzentrum\\52056 Aachen\\Germany}
\date{PITHA 97/45}
\maketitle
\begin{abstract}
We recalculate the hadronic contribution to the effective electromagnetic 
coupling constant $\alpha(M^2_Z)$ using  quasi-analytical approximations 
for the cross-section of the $e^+e^- \rightarrow hadrons$ proposed 
earlier by one of the authors (N.V.K., Mod.Phys.Lett.A9(1994)2825).   
We find that $\alpha^{-1}(M^2_Z) = 128.98 \pm 0.06$.
\end{abstract}

\newpage
For LEP1 observables at the Z-pole the largest effect from pure QED 
corrections is the change in the effective electromagnetic constant when 
going from $q^2=0$, where the fine structure constant $\alpha^{-1} = 
137.036$ is measured to $q^2=M^2_Z$ \cite{1}. The change from 
$\alpha \equiv \alpha(q^2=0)$ to $\alpha(M^2_Z)$ is related to the photon 
vacuum polarisation function $\Pi(q^2)$ via the relation
\begin{equation}
\alpha(q^2) = \frac{1}{1- \Pi(q^2)} ,
\end{equation}
where in the leading log approximation 
\begin{equation}
\Pi(q^2) = \Pi_{lept}(q^2) + \Pi_{had}(q^2),
\end{equation}
\begin{equation}
\Pi_{lept}(q^2)= \frac{\alpha}{3\pi}\sum_{i}(\ln(\frac{q^2}{m^2_i})-
\frac{5}{3} + O(\frac{m^2_i}{q^2}))
\end{equation}
and the hadronic contribution $\Pi_{had}(q^2)$ can be directly determined 
by the total cross-section of $e^+e^-$-annihilation into hadrons \cite{1}
\begin{equation}
Re (\Pi_{had}(s)) = -\frac{\alpha s}{3\pi} P \int_{4m^2_{\pi}}^{\infty}
\frac{R(s^{'})}{s^{'}(s^{'}-s)}ds^{'} ,
\end{equation}
\begin{equation}
\sigma_{h}(s) = \sigma_{tot}(e^+e^- \rightarrow \gamma 
\rightarrow  hadrons) =
\frac{4\pi \alpha^{2}}{3s}R(s)
\end{equation}
So the main problem in the calculation of $\alpha(M^2_Z)$ is the evaluation of 
the integral (4). There are several papers \cite{2} - \cite{9} devoted to 
the calculation of the integral (4). 

In this paper we recalculate the hadronic contribution to $\alpha(M^2_Z)$ 
using quasianalytical approximations for the cross-section of 
$e^+e^- \rightarrow hadrons$ proposed earlier by one of the authors \cite{4}.
The motivation for such recalculation is the appearance of 
new data on $e^{+}e^{-}$ - annihilation into hadrons in low energy region and 
new value  of the effective strong coupling constant $\alpha _{s}(M^2_Z)$. 
We find that 
\begin{equation}
\alpha^{-1}(M^2_Z) = 128.98 \pm 0.06
\end{equation}
In ref.\cite{4} the integral (4) has been calculated using quasianalytical 
approximation for $R(s)$. Namely, low energy contribution for $R(s)$ has 
been taken from experimental data and the theoretical ansatz has been used 
for high energy contribution for $R(s)$.  Two slightly different 
methods have been used. The first method numerically accounts for 
all resonances ($\rho,\omega, \phi, J/\psi, \Upsilon, ...$) and low energy 
continuum region $\sqrt{s} \leq 2.3$ GeV. The continuum contribution to 
$R(s)$ is determined by step-like approximation. In the approximation when 
strong coupling constant $\alpha_{s}=0$ our ansatz has the form \cite{4} 
\begin{equation}
R^{c}(s) = 2\theta(s-s_1) + \frac{4}{3}\theta(s-s_2) + 
\frac{1}{3} \theta(s-s_3) ,
\end{equation}
where we took $s_1 = (2.3 GeV)^2$, $s_2 = (4.4 GeV)^2$ and $s_3=(12 GeV)^2$. 
The parameters $s_2$ and $s_3$ correspond to the charm and beauty 
thresholds correspondingly. So in the first method we use the relation
\begin{equation}
\Pi_{had}(s) = \Pi_{2m_{\pi}-2.3 GeV}(s) + \Pi_{r, J/\psi,\Upsilon}(s) + 
\Pi_{c}(s)
\end{equation}
Here $\Pi_{2m_{\pi} - 2.3 GeV}(s)$ is the contribution of the low energy 
region $\sqrt{s} \leq 2.3 GeV$, $\Pi_{r, J/\psi, \Upsilon}$ is the 
contribution of $J/\psi$- , $\Upsilon$-resonances and their radial 
excitations and $\Pi_{c}(s)$ is the contribution of the continuum 
which is described by the formula (7) in zero approximation. 
In our paper we use the value \cite{5}
\begin{equation}
\Pi_{2m_{\pi}-2.3 GeV}(M^2_Z)= (6.06 \pm 0.25)\cdot 10^{-3}
\end{equation}
for the contribution of the low energy region. For the $J/\psi$-, 
$\Upsilon$-resonances and their radial excitations we use the formula 
\cite{3} 
\begin{equation}
\Pi_{res}(M^2_Z)  = \sum _{i} \frac{3\Gamma _{ee,i}}{M_i}\frac{\alpha}
{(\alpha(M^2_i))^2} ,
\end{equation}
where $M_i$ and $\Gamma_{ee,i}$ are the mass and leptonic width of the 
i-th resonance, respectively, and  the effective QED coupling constant 
at the resonance scale is used. An account of QCD and quark mass corrections 
lead to the appearance of the factors in formula (7) 
\begin{equation}
(1+\frac{2m^2_q}{s})\sqrt{(1-\frac{4m^2_q}{s})}[1+\frac{\alpha_s}{\pi}
f_1(\frac{m^2_q}{s}) +(\frac{\alpha_s}{\pi})^2  f_2(\frac{m^2_q}{s}) 
+(\frac{\alpha_s}{\pi})^3 f_3(\frac{m^2_q}{s})+...]
\end{equation}  
The coefficients $f_2(0)$ and $f_3(0)$ have been calculated in refs. 
 \cite{10} and the function $f_1(\frac{m^2_q}{s})$ is approximately determined 
by the expression 
\begin{equation}
f_1(x)= \frac{4\pi}{3}[\frac{\pi}{2v(x)}-\frac{3+v(x)}{4}(\frac{\pi}{2}-
\frac{3}{4\pi})],
\end{equation}
\begin{equation}
v(x)=\sqrt{(1-4x)}
\end{equation}

The second method of the calculations consists of taking into account low 
energy region $\sqrt{s} \leq 2.3$ GeV in the calculation of the integral (4) 
and an account of the high energy contribution to $\Pi_{had}(M^2_Z)$ 
is performed by taking into account nonzero c- and b-quark masses. So in the 
second method our ansatz has the form
\begin{equation}
\Pi_{had}(s) = \Pi_{2m_{\pi} -2.3 GeV}(s) + \Pi_{c}(s) ,
\end{equation} 
where in the approximation when strong coupling constant $\alpha_{s} =0$ 
our ansatz for $R_{cont}(s)$ is 
\begin{equation}
R_{cont}(s)= 2\theta(s-s_1) + \frac{4}{3}\theta(s-4m^2_c)(1 + 
\frac{2m^2_c}{s})\sqrt{(1-\frac{4m^2_c}{s})} + \frac{1}{3}\theta(s-4m^2_b)
(1+ \frac{2m^2_b}{s})\sqrt{(1- \frac{4m^2_b}{s})}
\end{equation}
Nonzero c- and b-quark masses in formula (15) effectively take into account 
the contribution of $J/\psi$- and $\Upsilon$-resonances and their radial 
excitations. Such approach works rather well in the method of QCD sum rules.

In our numerical calculations we used the value of the effective 
strong coupling constant equal to \cite{11}
\begin{equation}
\alpha_s(M^2_Z) = 0.119  \pm 0.006
\end{equation}
and the values of the c- and the b- pole quark masses equal to \cite{12} 
\begin{equation}
m_c^{pole} = (1 +\frac{4}{3} \frac{\alpha_{s}(m_c(m_c))}{\pi} +...)
m_c(m_c) = 1.4\cdot(1 \pm 0.1) GeV ,
\end{equation}
\begin{equation}
m_b^{pole} = (1 + \frac{4}{3}\frac{\alpha_{s}(m_b(m_b))}{\pi} +...)
m_b(m_b) = 4.5 \cdot (1 \pm 0.1) GeV
\end{equation}

In the method 1 the inverse effective electromagnetic 
coupling constant $\alpha^{-1}(M^2_Z)$ is represented in the form
\begin{equation}
\alpha^{-1}(M^2_Z) = \alpha^{-1} - \Delta \alpha^{-1}(l) - 
\Delta \alpha^{-1}(\sqrt{s} < 2.3 GeV) -\Delta \alpha^{-1}(J/\psi, \Upsilon) 
-\Delta \alpha^{-1}(c) -\Delta \alpha^{-1}(t)
\end{equation}
Here $\Delta \alpha^{-1}(l)$ is the leptons contribution, 
$\Delta \alpha^{-1}(\sqrt{s} < 2.3 GeV)$ is the low energy contribution, 
$\Delta \alpha^{-1}(J/\psi,\Upsilon)$ is the contribution of $J/\psi$-,
$\Upsilon$-resonances and their radial excitations, $\Delta \alpha^{-1}(c)
$ is the continuum contribution and $\Delta \alpha^{-1}(t)$ is 
the top-quark contribution. In the estimation of $\Delta \alpha^{-1}(l)$ 
we have used two loop approximation. To estimate the uncertainties  we 
assumed the uncertainties related with the choise of $s_2$ and $s_3$ to be 
equal 20 pecent. In our calculations we used the values $f_2(\frac{m^2_q}{s})$ 
and $f_3(\frac{m^2_q}{s})$ at $m^2_q=0$. However the dependence of our results 
in method 1 on the value of c- and b-quark masses is rather small and also 
three and four loop contributions are also small numerically. As it has been 
mentioned before in the estimation of low energy hadron contribution into 
$\alpha^{-1}(M^2_Z)$ we used the results of ref. \cite{5}. Numerically 
we have found \footnote{In the calculation of $\Delta \alpha^{-1}(J/\psi, 
\Upsilon)$ we have used the data from ref. \cite{13}}  
\begin{equation}
\Delta \alpha^{-1}(l) = 4.313 ,
\end{equation}
\begin{equation}
\Delta \alpha^{-1}(\sqrt{s} < 2.3 GeV) = 0.830 \pm 0.034 ,
\end{equation}
\begin{equation}
\Delta \alpha^{-1}(J/\psi, \Upsilon) =0.160 \pm 0.008(0.016),
\end{equation}
\begin{equation}
\Delta \alpha^{-1}(c) = 2.757 \pm 0.019(\alpha_s) \pm 0.024(h.c.)
\pm 0.003(m_c,m_b) \pm 0.032(s_2) \pm 0.008(s_3),
\end{equation}
\begin{equation}
\Delta \alpha^{-1}(m_t=174 GeV) = - \frac{4}{3\pi} \cdot \frac{1}{15} 
\frac{M^2_Z}{m^2_t} \approx -0.008
\end{equation}
In formula (23) the uncertainties $0.019(\alpha_s)$, $0.003(m_c, m_b)$, 
$0.032(s_2)$, $0.008(s_3)$ are determined by the uncertainties in the 
determination of $\alpha_{s}(M^2_Z)$, c- and b-quark masses, parameters of the 
spectrum $s_2$ and $s_3$. The uncertainty $0.024(h.c.)$ is our 
estimate of the higher order contributions. We assume that such uncertainty 
is equal to one halph of the difference between calculated value which 
takes into account QCD corrections up to 3 loops and one loop 
contribution or numerically it coincides with 3 loop contribution. 
Note that we assumed 20 percent uncertainty in the determination of 
$s_2$ and $s_3$ that is rather conservative estimate. In formula (22) in 
the estimation of the overall error we assumed the errors in the 
determination of the contribution of different resonances are 
statistically independent and 
obtained the error equal to 0.008. The number in brackets corresponds to 
the case when we simply sum up the errors from different resonances. 
Assuming that all errors are statistically independent we find 
\begin{equation}
\alpha^{-1}(M^2_Z) = 128.98 \pm 0.06(0.13)
\end{equation}
In formula (25) the number in brackets corresponds to the case when we 
simply sum up the errors. 

In method 2 the inverse effective electromagnetic coupling constant 
$\alpha^{-1}(M^2_Z)$ can be represented in the form 
\begin{equation}
\alpha^{-1}(M^2_Z)=\alpha^{-1} -\Delta \alpha^{-1}(l)- \Delta 
\alpha^{-1}(\sqrt{s} <2.3 GeV) - \Delta \alpha^{-1}(c)-
\Delta \alpha^{-1}(t)
\end{equation}
Here $\Delta \alpha^{-1}(c)$ is the contribution of heavy quark 
continuum and heavy quark resonances. Numerically we have found 
\begin{equation}
\Delta \alpha^{-1}(c) = 2.950 \pm 0.033(\alpha _{s}) 
\pm 0.038(h.c.) 
\pm 0.030(m_{c}) \pm 0.009(m_{b})
\end{equation}
Here uncertainties $0.033(\alpha_s)$, $0.038(h. c.)$, $0.0030(m_c)$, 
$0.008(m_b)$ are the uncertainties determined by the uncertainties of 
$\alpha _{s}(M^2_Z)$, higher order corrections, c-quark mass an b-quark mass. 
Assuming that all errors are statistically independent we find that 
\begin{equation}
\alpha^{-1}(M^2_Z) = 129.95 \pm 0.07(0.14)
\end{equation}
in the method 2. The number in brackets in the formula (28) corresponds to the 
simple summation of the errors. So we have found that both methods 1 and 2 
lead to similar values for  $\alpha^{-1}(M^2_Z)$ with the similar errors.
However we believe that method 1 is more reliable since the errors related 
with the $\alpha_{s}(M^2_Z)$ uncertainty and the uncertainty of higher order 
corrections are smaller in method 1. Besides, in the estimation of the errors 
in method 1 we assumed very conservative estimates in the uncertainties 
related with the choise of $s_2$ and $s_3$ (20 percent). For instance, 
if we assume 10 percent uncertainty in  the choise of $s_2$ and $s_3$ 
our error in formula (28) will be 0.05(0.10). Therefore we quote in the 
abstract our estimate of $\alpha^{-1}(M^2_Z)$ obtained using the method 1. 
The value obtained in our paper is very similar to the value 
$\alpha^{-1}(M^2_Z) = 128.97 \pm 0.06(exp.) \pm0.07(theor.)$ obtained 
in ref.\cite{4}. The decrease of the errors is related mainly with the better 
calculation of low energy contribution $\Delta  \alpha^{-1}(M^2_Z)(\sqrt{s} 
< 2.3 GeV)$ and the better determination of the $\alpha_{s}(M^2_Z)$. 
As it has been mentioned before there are several recent calculations of 
$\alpha^{-1}(M^2_Z)$. Table 1 shows a comparison of some recent estimates 
of $\alpha^{-1}(M^2_Z)$. 

To conclude, in this paper we have recalculated the value of 
$\alpha^{-1}(M^2_Z)$ using two methods of ref.\cite{4} 
and new experimental data. We have found that both methods give similar 
results, however as it has been mentioned before we believe that 
the first method  is more reliable.

We are indebted to RFFI-DFG research program project No. 436 RUS 
113/227/0 which made possible our collaboration. 

\newpage

Table 1. Some recent values of $\alpha^{-1}(M^2_Z)$ 

\begin{center}
\begin{tabular}{|l| |l| }
\hline
$\alpha^{-1}(M^2_Z)$ & ref. \\
\hline
$ 128.87  \pm 0.12$  & [3] \\
\hline
$128.97 \pm0.06(exp.)\pm 0.07(theor.)$& [4]\\
\hline0
$128.99 \pm 0.06$ & [5]\\
\hline
$128.89 \pm0.06$ & [6]\\
\hline
$128.96 \pm 0.06$& [7]\\
\hline
$128.896 \pm 0.090$& [8]\\
\hline
$128.98 \pm 0.06$& this paper\\
\hline
\end{tabular}
\end{center}

\end{document}